\documentclass[aps,prx,twocolumn,superscriptaddress,noshowpacs]{revtex4-2}
\usepackage{graphicx}
\usepackage{xcolor}
\usepackage[normalem]{ulem}
\usepackage{float}
\usepackage{braket}
\usepackage{placeins}
\usepackage{amsmath}
\usepackage{physics}
\usepackage{lipsum}
\usepackage{verbatim}
\usepackage{float}
\usepackage{ulem}
\usepackage{times}
\usepackage{soul,color}
\usepackage{bm}
\usepackage{amsmath,amssymb}
\usepackage[dvipsnames]{xcolor}
\usepackage{mhchem}
\usepackage{url}
\usepackage[normalem]{ulem}
\usepackage{bm,nicefrac,xfrac}
\usepackage{xspace}
\usepackage{enumitem}

\begin{document}

\title{Translational symmetry breaking in the electronic nematic phase of BaFe$_2$As$_2$}

\author{K.~Koshiishi}
\affiliation{Department of Physics, The University of Tokyo, Bunkyo-ku, Tokyo 113-0033, Japan}

\author{L.~Liu}
\affiliation{Department of Physics, The University of Tokyo, Bunkyo-ku, Tokyo 113-0033, Japan}

\author{K.~Okazaki}
\affiliation{Department of Physics, The University of Tokyo, Bunkyo-ku, Tokyo 113-0033, Japan}
\affiliation{Institute for Solid State Physics, The University of Tokyo, Kashiwa,
Chiba 277-8581, Japan}

\author{H.~Suzuki}
\affiliation{Department of Physics, The University of Tokyo, Bunkyo-ku, Tokyo 113-0033, Japan}
\affiliation{Institute of Multidisciplinary Research for Advanced Materials (IMRAM), Tohoku University, Sendai 980-8577, Japan}
\affiliation{Frontier Research Institute for Interdisciplinary Sciences, Tohoku University, Sendai 980-8578, Japan}

\author{J.~Xu}
\affiliation{Department of Physics, The University of Tokyo, Bunkyo-ku, Tokyo 113-0033, Japan}

\author{M.~Horio}
\affiliation{Department of Physics, The University of Tokyo, Bunkyo-ku, Tokyo 113-0033, Japan}
\affiliation{Institute for Solid State Physics, The University of Tokyo, Kashiwa,
Chiba 277-8581, Japan}

\author{H.~Kumigashira}
\affiliation{KEK, Photon Factory, Tsukuba, Ibaraki 305-0801, Japan}
\affiliation{Institute of Multidisciplinary Research for Advanced Materials (IMRAM), Tohoku University, Sendai 980-8577, Japan}

\author{K.~Ono}
\affiliation{KEK, Photon Factory, Tsukuba, Ibaraki 305-0801, Japan}
\affiliation{Department of Applied Physics, Osaka University, Suita 565-0871, Osaka, Japan}

\author{M.~Nakajima}
\affiliation{National Institute of Advanced Industrial Science and Technology, Tsukuba, Ibaraki, 305-8568, Japan}
\affiliation{RIKEN Center for Emergent Matter Science, Wako, Saitama 351-0198, Japan}
\affiliation{Department of Physics, Osaka University, Toyonaka, Osaka 560-0043, Japan}

\author{S.~Ishida}
\affiliation{National Institute of Advanced Industrial Science and Technology, Tsukuba, Ibaraki, 305-8568, Japan}

\author{K.~Kihou}
\affiliation{National Institute of Advanced Industrial Science and Technology, Tsukuba, Ibaraki, 305-8568, Japan}

\author{C.~H.~Lee}
\affiliation{National Institute of Advanced Industrial Science and Technology, Tsukuba, Ibaraki, 305-8568, Japan}

\author{A.~Iyo}
\affiliation{National Institute of Advanced Industrial Science and Technology, Tsukuba, Ibaraki, 305-8568, Japan}

\author{H.~Eisaki}
\affiliation{National Institute of Advanced Industrial Science and Technology, Tsukuba, Ibaraki, 305-8568, Japan}

\author{S.~Uchida}
\affiliation{Department of Physics, The University of Tokyo, Bunkyo-ku, Tokyo 113-0033, Japan}
\affiliation{National Institute of Advanced Industrial Science and Technology, Tsukuba, Ibaraki, 305-8568, Japan}

\author{A.~Fujimori}
\email{fujimori@phys.s.u-tokyo.ac.jp}
\affiliation{Department of Physics, The University of Tokyo, Bunkyo-ku, Tokyo 113-0033, Japan}
\affiliation{Department of Physics, National Tsing Hua University, Hsinchu 300044, Taiwan}

\begin{abstract}

The microscopic origin of the nematicity, namely, four-fold rotational symmetry breaking in iron-based superconductors has been controversial since its discovery. 
In particular, its relationship with the stripe-type spin-density-wave order and the orthorhombic lattice distortion in the antiferromagnetic orthorhombic (AFO) phase, which exists at temperatures below the electronic nematic phase, has been highly debated.  
Here, we report on the temperature evolution of angle-resolved photoemission spectra of the parent compound BaFe$_2$As$_2$, ranging from the AFO to nematic to paramagnetic phases.
The Dirac cone feature, which is formed in the AFO phase, is found to persist in the nematic phase, suggesting that an antiferroic order of the same periodicity as the AFO phase persists in the nematic phase. 
Considering the relatively shallow $d_{xy}$ orbital in BaFe$_2$As$_2$, 
we propose that an antiferro-orbital order 
involving the $d_{xy}$ and other orbitals
takes place in the nematic phase.

\end{abstract}
\date{\today}
\maketitle

\section{Introduction}

\begin{figure}[thp]
\centering
\includegraphics[width=1.0\columnwidth]{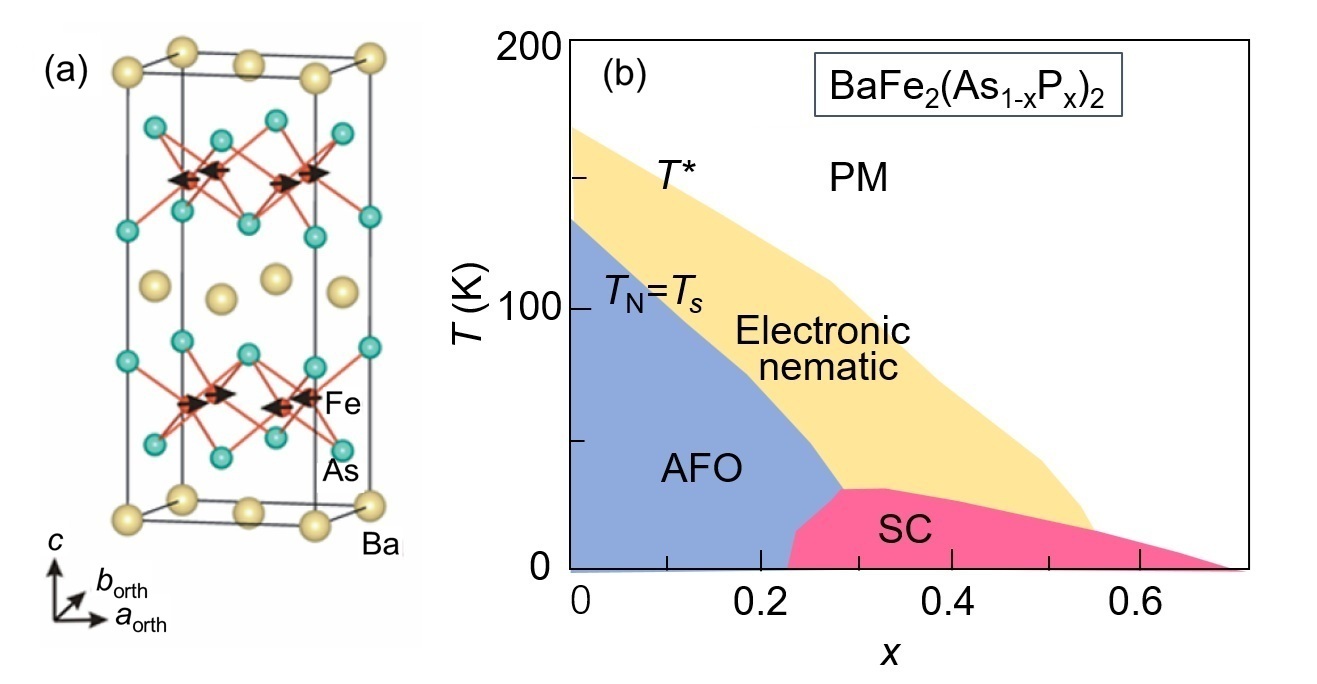}
\caption{(a) Crystal and magnetic structures of the 122-type BaFe$_2$As$_2$ in the antiferromagnetic orthorhombic (AFO) phase (adopted from Ref.~\onlinecite{qhuang}). (b) Phase diagram of the P-substituted BaFe$_2$As$_2$~\cite{kasahara}. PM: Paramagnetic phase; SC: Superconducting phase; $T_N$: N\'{e}el temperature: $T_s$: Tetragonal-to-orthorhombic transition temperature; $T^\ast$: Nematic transition temperature.}
\label{fig1}
\end{figure}

The parent compounds of a majority of iron-based superconductors (FeSCs) are antiferromagnetic (AFM) metals with a stripe-type spin-density wave (SDW) order accompanied by a tetragonal-to-orthorhombic lattice distortion. 
Figure~\ref{fig1}(a) shows the crystal and magnetic structures of BaFe$_2$As$_2$, one of the parent compounds of the so-called 122-type FeSCs.  
Below $T_{\rm N} = T_s$ ($T_{\rm N}$: N\'{e}el temperature; $T_s$: structural transition temperature), the system enters the antiferromagnetic-orthorhombic (AFO) phase~\cite{qhuang}.  
Carrier doping or chemical pressure by element substitution for BaFe$_2$As$_2$ (e.g., K for Ba; Co, Ni, or Cu for Fe; As for P) suppresses the AFO order and superconductivity emerges~\cite{sefat,ljli,canfield,sjiang,nni}, as shown in the phase diagram [Fig.~\ref{fig1}(b)]. Above the AFO and the superconducting (SC) phases up to the temperature denoted by $T^\ast$ in the figure, there appears the so-called electronic nematic phase~\cite{kasahara}. In that phase,  although the crystal lattice recovers the tetragonal symmetry, the electronic system still exhibits the broken four-fold rotational symmetry ($C_4$ symmetry) but shows a two-fold rotational one ($C_2$ symmetry).  
  
In the AFO phase, the SDW has the magnetic ordering wave vector of $\textbf{Q} = (\frac{1}{2},0,\frac{1}{2})$ in r.l.u. and the magnetic moments at the Fe sites are aligned ferromagnetically along the orthorhombic $b$ axis and antiferromagnetically along the orthorhombic $a$ axis and along the $c$ axis. 
The lattice constant along the $b$-axis is slightly smaller than that along the $a$-axis. 
In the paramagnetic (PM) phase, the tetragonal $I4/mmm$ symmetry is recovered and the body-centered-tetragonal (bct) unit cell contains two Fe atoms. In the stripe-type SDW phase below $T_{\rm N}=T_s\sim$ 140 K, the unit cell of orthorhombic $Fmmm$ symmetry [Fig.~\ref{fig1} (a)] contains four Fe atoms. 
Correspondingly, the Brillouin-zone of the bct lattice is folded below $T_{\rm N}=T_s$, as shown in Fig.~\ref{fig2}(c). In fact, the Fermi surface reconstruction [Fig.~\ref{fig2}(e)] and the folded band structure shown in have been identified by angle-resolved photoemission spectroscopy (ARPES) studies of BaFe$_2$As$_2$~\cite{myiPRX,kondo,lliu}. 

It has been known that, in the 122-type compounds, twinned domains are formed below the structural phase transition temperature $T_s$, but by applying uniaxial pressure, one can avoid the twinning~\cite{jhchu}. In the AFO phase, detwinned single crystals of the 122-type FeSCs have revealed the intrinsic in-plane anisotropy of the electronic structure detected by transport~\cite{jhchu,ishidaPRB}, optical conductivity~\cite{nakajima,dusza}, neutron scattering~\cite{jzhao}, and ARPES~\cite{lliu,myiPNAS} measurements. Furthermore, in some of these measurements, the same in-plane anisotropies continue up to a few to several tens degree above $T_s$~\cite{jhchu,ishidaPRB,myiPNAS,ishidaPRL,ishidaJACS}. Such behaviors where the electronic states exhibit spontaneously broken four-fold rotational symmetry are called electronic nematicity in analogy to an ordered phase of liquid crystals.
Evidence for the electronic nematicity above the structural transition temperature $T_s$ has also been found in single domain samples of BaFe$_2$(As$_{1-x}$P$_x$)$_2$ through magnetic torque measurements, as shown in Fig.~\ref{fig1}(b)~\cite{kasahara}. 
This anomaly is seen up to $ \sim T^\ast$ through the finite amplitude of two-fold symmetry signals in magnetic torque. 
In the ARPES study of detwinned Ba(Fe$_{0.975}$Co$_{0.025}$)$_2$As$_2$~\cite{myiPNAS}, too, it has been reported that the in-plane anisotropy of the band structure persists above $T_s$. 
Nematicity above $T_s$ has also been observed in transport and optical measurements~\cite{ishidaPRB,dusza}. 
In other strongly correlated electron systems such as cuprate high-temperature superconductors~\cite{achker,ishida-shibauchi,nakata} and heavy fermion systems~\cite{ronning}, too, nematicity has been reported. 
Recently, the electronic nematicity in the 11-type FeSCs such as FeSe~\cite{shimojima} is more extensively studied because they do not show AFM order and have simpler phase diagrams. 

Since FeSCs are multi-band systems, the orbital degrees of freedom is considered to play an important role.
Theoretically, a ferro-orbital order is proposed as an origin of the broken $C_4$-symmetry in iron pnictides~\cite{cclee}. 
On the other hand, no anomalies in the elastic constant $C_{66}$ have been observed at $T^\ast$ in ultrasonic measurements~\cite{yoshizawa}, implying that no transition to ferro-orbital order occurs at $T^\ast$. 
In this regard, an antiferro-orbital order has been proposed by Kontani \textit{et al.}~\cite{kontani}. Antiferro-quadrupole fluctuations induced by electron-phonon interaction may lead to an antiferro-orbital order in the electronic nematic phase of BaFe$_2$As$_2$ that can hardly be detected by long-wavelength (\textbf{q} $\simeq$ 0) probes such as the $C_{66}$ measurements.

In the present paper, we report on an ARPES study of detwinned BaFe$_2$As$_2$ crystals to reveal the temperature evolution of band-folding features from the AFO phase to the electronic nematic to paramagnetic (PM) tetragonal phases. 
We find that a Dirac-cone feature near the X point of the original Brillouin zone, which is created by the AFM band folding due to the SDW order, persists above $T_{\rm N} \simeq T_s =$ 142 K up to $T^\ast\sim$ 170 K. Our results indicate that the band folding persists in the electronic nematic phase even after the AFM order vanishes above $T_s=T_{\rm N}$, suggesting that the nematic phase has the same periodicity as the stripe-type SDW order. 
Although the nature of the ordering cannot be uniquely determined by ARPES alone, the present results demonstrate that an antiferro-ordered phase persists above $T_s$ up to $T^\ast$ in BaFe$_2$As$_2$.

\begin{figure*}[tbp]
\centering
\includegraphics[width=1.7\columnwidth]{Figure2.jpg}
\caption{ARPES spectra of BaFe$_2$As$_2$ in the antiferromagnetic-orthorhombic phase. (a), (b) Fermi surface mapping for the detwinned and twinned BaFe$_2$As$_2$ crystals, respectively. Inset to (a) shows the crystal and magnetic structure on the 2D plane and the direction of the applied strain P. (c) 3D Brillouin zones of the PM (gray) and AFO (green) phases. (d) Corresponding in-plane 2D Brillouin zones. (e) Schematic Fermi surfaces. Orange and blue circles indicate hole pockets and electron pockets, respectively. (f), (h) Energy-momentum plots along cuts 1 and 2 indicated by red arrows in (a). (g), (i) Second-derivative plots of EDCs corresponding to cut 1 and cut 2.
}
\label{fig2}
\end{figure*}

\section{methods}

High-quality single crystals of BaFe$_2$As$_2$ ($T_{\rm N} \simeq T_s =$ 142 K) were grown by the self-flux method followed by annealing~\cite{ishidaPRB,nakajima,ishidaPRL}. 
In order to detect the anisotropic electronic structure by ARPES measurements, we detwinned the single crystals of BaFe$_2$As$_2$ by applying uniaxial strain using a specially designed sample holders similar to that used in Ref.~\onlinecite{ykim}. 
ARPES measurements were performed at beamline 28A of Photon Factory, High-Energy Accelerator Research Organization (KEK) using a SCIENTA SES-2002 electron analyzer. 
The total energy resolution was set to $\sim$ 20 meV. 
ARPES spectra were obtained using circularly-polarized light with the photon energy of 63 eV, corresponding to $k_z\sim 2\pi/c$. 
The crystals were cleaved \textit{in situ} at 20 K for low-temperature measurements and at 100 K for temperature-dependent measurements. 
The measurements were carried out in an ultrahigh vacuum of $\sim 4\times 10^{-10}$ Torr. 
The Fermi level ($E_{\rm F}$) of the samples was calibrated using gold. 

Density-functional-theory (DFT) calculations using the full-potential linearized augmented plane-wave (FP-LAPW) method were performed with a WIEN2k package~\cite{blaha}. 
The effective potential is calculated using the generalized gradient approximation (GGA-PBE). 
The experimental lattice parameters of Ba$_{0.6}$K$_{0.4}$Fe$_2$As$_2$ with the tetragonal $I4/mmm$ structure~\cite{rotter} were used.

\section{results and discussion}

First, in order to confirm whether the samples were sufficiently detwinned or not, we performed ARPES measurements on both twinned and detwinned samples in the AFO phase. 
Figures~\ref{fig2}(a) and \ref{fig2}(b) show Fermi surface (FS) mapping in the $k_x$-$k_y$ planes of the detwinned and twinned samples, respectively. 
The geometry of the FSs of the twinned sample [panel (b)] approximately show four-folded rotational symmetry around the Z point and the ZX and ZY directions are similar. 
On the other hand, for the detwinned BaFe$_2$As$_2$ [panel (a)], differences of the FSs around the Z-X and Z-Y lines can be clearly seen as already reported in Ref.~\onlinecite{lliu}, i.e., the FSs of the detwinned sample show two-fold rotational symmetry. 
For example, bright small spots around the Y point observed in the twinned sample disappear in the detwinned sample, which is consistent with the previous ARPES study~\cite{myiPNAS}. 

In Fig.~\ref{fig2}(c), the three-dimensional Brillouin zone (BZ) of the PM state and the AFM state are depicted by gray and green lines, respectively. 
In the PM phase, the BZ has four-fold rotational symmetry around the $\Gamma$-Z line and the X and Y points are equivalent. 
In the AFO phase, because of the stripe-type order of the magnetic moments at the Fe sites, the unit cell is doubled and contains four Fe atoms. 
Hence, the energy bands and FSs of the PM phase are folded into the AFM BZ of half volume. In addition, the Z($\Gamma$) point should be equivalent to the Y(X) point but non-equivalent to the X(Y) point in the AFM BZ as shown in Fig.~\ref{fig2}(d). 
In Fig.~\ref{fig2}(a), FSs around the Z point are similar to those around the Y point while they are different from those around the X point, suggesting that FSs intrinsic to the AFO phase were observed and hence that the crystals were properly detwinned.

Focusing on the geometry of the FSs in the AFO phase, we found that they consist of a circular hole pocket centered at the center of the AFM BZ, two elliptical electron pockets next to the hole pocket on the Z-Y line, and two high-intensity tiny electron pockets on the Z-X line, which correspond to the apex of the Dirac cone as reported in Ref.~\onlinecite{richard}, denoted as $\alpha/\beta$, $\epsilon$, and $\delta$, respectively. 
A schematic figure of the FSs in the AFO phase we have observed are shown in Fig.~\ref{fig2}(e). 
The orange and blue circles correspond to hole and electron Fermi surfaces, respectively. 
Terashima \textit{et al.}~\cite{terashima} have also determined the Fermi surfaces similar to our results by Shubnikov-de Hass oscillation measurements
~\footnote{One may point out that the hole Fermi surface could not be observed at the Z point by the Shubnikov-de Haas oscillation measurements. However, we consider that the photon energy of 63 eV used in our measurements did not exactly correspond to the Z point, or that $k_z$-integrated FSs was observed due to momentum resoluion. In spite of that, our results are basically the same as the previous study.}.
Such elliptical electron pockets and tiny electron pockets should not be seen in the PM phase, where three hole FSs around the $\Gamma$ point and two electron FSs around the zone corner exist. 
Thus, the observed FS features are characteristics of the AFO phase with the Fermi-surface reconstruction caused by the stripe-type SDW. 

Binding energy ($E_B$) \textit{versus} momentum (\textbf{k}) plots across the hole and electron pockets along cuts 1 and 2 [red arrows in Fig.~\ref{fig2}(a)] are shown in Figs.~\ref{fig2}(f) and \ref{fig2}(h), respectively. 
Figures~\ref{fig2}(g) and \ref{fig2}(i) show the second-derivative plots of these spectra with respect to energy. For cut 1, two shallow electron bands are seen near $E_\text{F}$. For cut 2, one can see two linear bands intersecting each other near $E_\text{F}$, which results from the band folding due to the SDW order. 

\begin{figure}[thp]
\centering
\includegraphics[width=1.0\columnwidth]{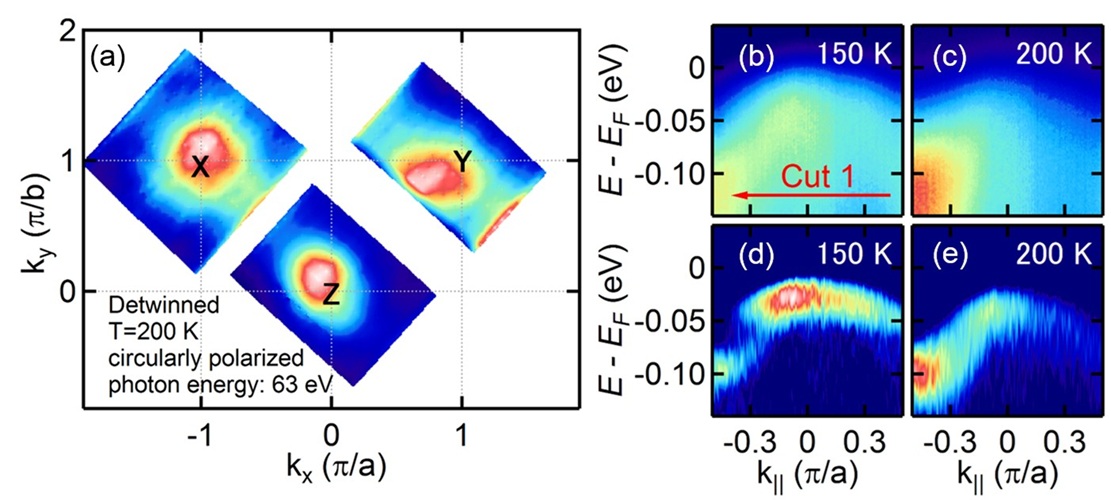}
\caption{ARPES spectra of BaFe$_2$As$_2$ in the paramagnetic (PM) tetragonal phase. (a) Fermi-surface mapping. The spectra were taken at $T=$ 200 K. (b), (c) Energy-momentum plots along cut 1 in Fig. 1(a) for $T=$ 150 K and 200 K. (d), (e) Second-derivative plots of (b) and (c) with respect to energy.}
\label{fig3}
\end{figure}

ARPES spectra observed in the PM phase well above $T_s =$ 142 K are shown in Fig.~\ref{fig3}. Figure~\ref{fig3}(a) shows in-plane FS mapping at 200 K. 
The FSs clearly changed in shape: $\delta$ electron pockets near the X point disappeared and electron pockets appeared at the X and Y points. ARPES spectra along cut 1 for $T=$ 150 K and 200 K are shown in Figs.~\ref{fig3}(b) and \ref{fig3}(c), respectively. 
Figures~\ref{fig3}(d) and \ref{fig3}(e) show their second-derivative plots. 
In comparison with the spectra in the AFO phase [Figs.~\ref{fig2}(f) and \ref{fig2}(g)], the shallow electron-like bands near $E_{\rm F}$ became increasingly difficult to observe with temperature and that only hole bands centered at the Z point remained at 200 K.

\begin{figure}[tbp]
\centering
\includegraphics[width=1.0\columnwidth]{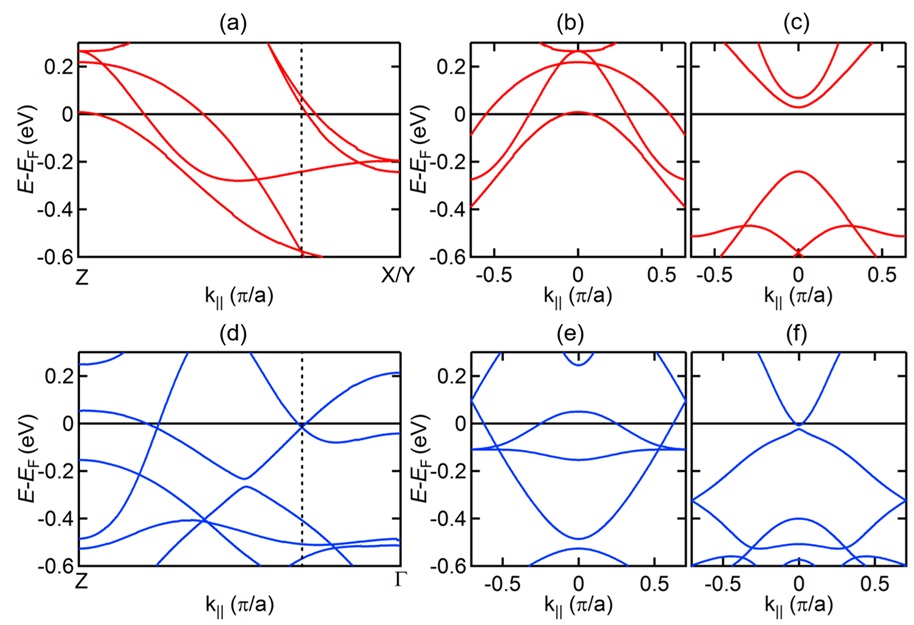}
\caption{DFT-GGA calculation of the ban structures of BaFe$_2$As$_2$. (a) (d) Band structures of the PM and AFM states, respectively, along the Z-X lines. 
The Z-X line is perpendicular to cuts 1 and 2. (b), (e) Calculated band structure of the PM and AFM states, respectively, along cut 1. The center of cuts correspond to the Z point and cuts are perpendicular to the Z-X line. 
(c), (f) Calculated results for cut 2 in the PM and AFM states, respectively. 
Cut 2 is parallel to cut 1 and crosses the Dirac point indicated by the broken line in (a) and (d).} 
\label{fig4}
\end{figure}

\begin{figure}[tbp]
\centering
\includegraphics[width=1.0\columnwidth]{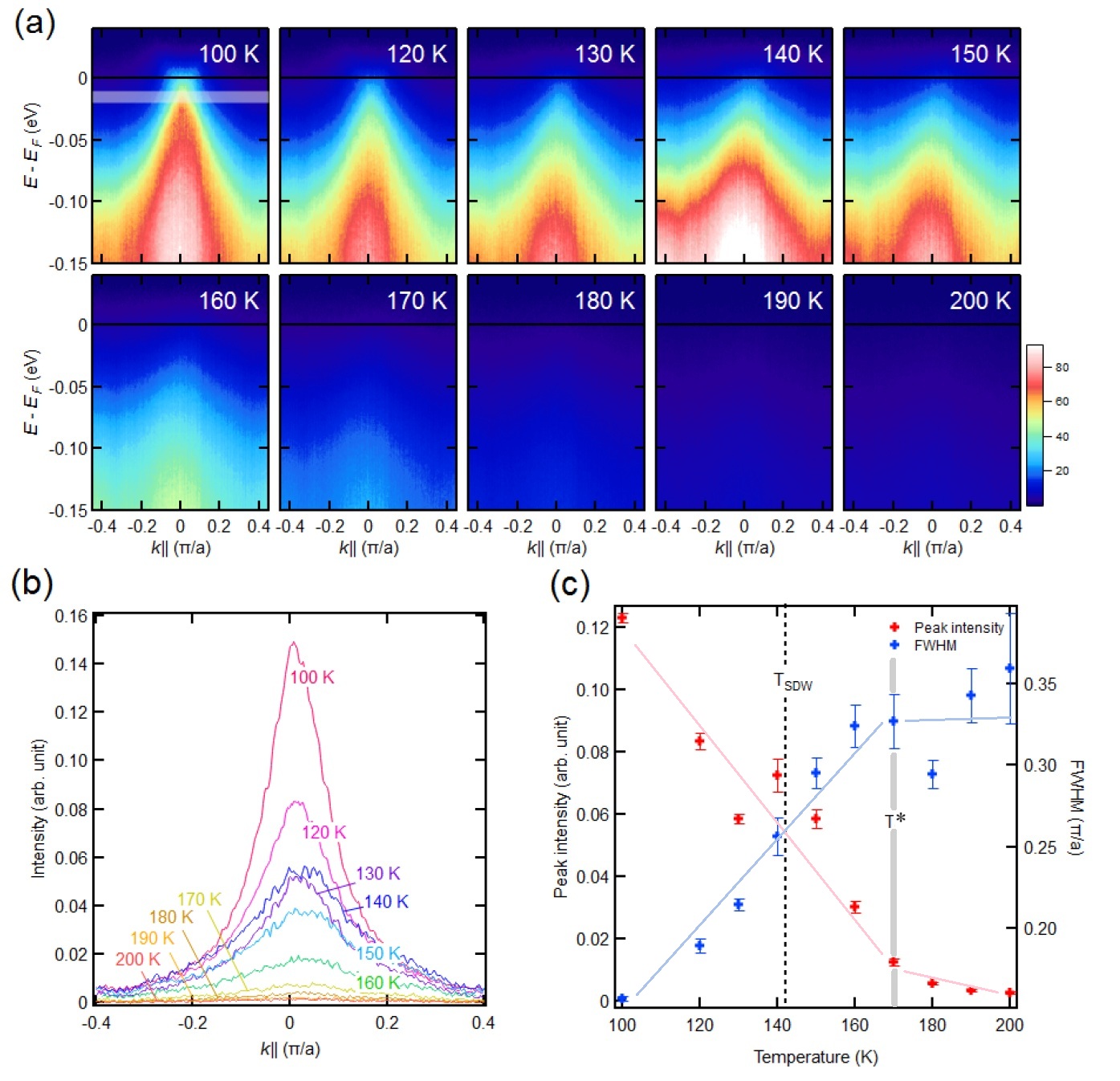}
\caption{Temperature dependence of the Dirac-cone feature in ARPES spectra. (a) Evolution of ARPES spectra of BaFe$_2$As$_2$ with temperature. Spectra from 100 K to 200 K are taken along cut 2 shown in Fig.~\ref{fig2}(a). The spectra have been divided by the FermiDirac function. (b) Temperature evolution of the momentum distribution curve (MDC) for the binding energies form 10 meV to 20 meV. (c) Temperature dependence of the MDC peak intensity and width. }
\label{fig5}
\end{figure}

In order to see predicted changes in the band dispersions between the AFO phase and the PM phase, we have performed DFT calculations on BaFe$_2$As$_2$ in the PM and AFM states, as shown in Fig.~\ref{fig4}. 
Red and blue band dispersions are those in the PM and AFM states, respectively. 
Figures \ref{fig4}(a) and (d) show calculated band structures along the Z-X($\Gamma$) direction, Figs.~\ref{fig4}(b) and (e) cut 1, and Figs.~\ref{fig4}(c) and (f) cut 2. 
Cut 2 shown in Figs.~\ref{fig4}(c) and \ref{fig4}(f) crosses the Z-X($\Gamma$) line at the dotted line positions shown in Figs.~\ref{fig4}(a) and \ref{fig4}(d). 
Note that the $\Gamma$ point in the AFM BZ is equivalent to the X/Y point in the PM BZ. In the PM state, one can see three hole bands centered at the Z point and two electron bands centered at the X/Y point [Figs.~\ref{fig4}(a) and (b)]. 
In the AFM state, original bands are hybridized with folded bands and form a complicated band structure [Figs.~\ref{fig4}(d)-(f)]. 
One can see two crossing points near $E_{\rm F}$ in Fig.~\ref{fig4}(d), which are formed by an original hole band and a folded electron band. 
Since the parities of the two bands are different on the high-symmetry line, they cannot hybridize with each other there. 
Indeed, in Figs.~\ref{fig4}(d) and (f), it is clear that two bands cross each other near $E_{\rm F}$. 
This means that a Dirac cone should be formed near the Fermi level due to the band folding. 
The calculated band structure for the AFM state is in agreement with the Dirac-cone-like band dispersion observed in our ARPES measurements, indicating that the band structure observed in the present experiment is the intrinsic electronic structure of the AFO phase.

To investigate the electronic structure in the electronic nematic phase, 
we have performed temperature-dependent measurement of the Dirac-cone band at various temperatures across $T_s =$ 142 K and $T^\ast\sim$ 170 K. 
Figure~\ref{fig5}(a) shows $E_B$ \textit{versus} $\textbf{k}$ plots around the apex of the Dirac cone between $T=$ 100 K and 200 K. The linear band crossing can be seen at 100 K as in the case of 20 K. 
As the temperature increases, the Dirac cone is gradually weakened but persists above $T_{\rm N} \simeq T_s =$ 142 K, and eventually vanishes around $T^\ast\sim$ 170 K. 
For quantitative comparison, momentum distribution curve (MDC) integrated within the binding energies from 10 meV to 20 meV [Fig.~\ref{fig5}(a)] are shown in Fig.~\ref{fig5}(b). 
The figure clearly shows that the Dirac cone gradually weakens upon heating but persists above $T_s$. 
Figure~\ref{fig5}(c) shows the temperature dependence of the peak intensity obtained by integrating the MDC peak from 10 to 20 meV below $E_{\rm F}$. 
As the temperature increases, the peak intensity gradually decreases. We have also plotted the full width at half maximum (FWHM) of the MDC peak in Fig.~\ref{fig5}(c). The width significantly increases around $T_s$ upon heating and saturates around $\sim$ 160 K, suggesting that change of the electronic structure associated with the phase transition persists well above $T_s$. The weak strain by the detwinning is known to induce finite nematicity up to $\sim$160 K~\cite{ren-yli}, but cannot explain the band-folding effect observed in the present study. 

The present results suggest that short-range SDW or dynamical SDW of the stripe type persists up to $\sim T^{\ast}$ or that some electronic order (or short-range order) which has the same periodicity as the stripe-type SDW order remains above $T_{\rm N} = T_S$. 
As for the possibility of short-range or dynamical SDW, only nematic spin correlations, i.e., the elongation of Bragg spots along one direction, has been reported above $T_{\rm N}=T_S$ by inelastic neutron scattering studies of detwinned crystals~\cite{xlu,man-dai}. 
As for the orbital degrees of freedom of the electronic structure, $\pm O_{xz}$-antiferro-quadrupole order ($xy\pm xz$ orbital order) has been theoretically proposed as a possible antiferro-orbital order in the 122-type Fe pnictides (because of the proximity of the Fe $xy$ orbital to the Fermi level compared to the 11-type ones)~\cite{kontani}. 
This orbital order has the same spatial periodicity as the stripe-type SDW. 
In Fe-based superconductors, it has been considered that the orbital degrees of freedom play an important role because the six electrons of Fe$^{2+}$ occupy five $3d$ orbitals located near the Fermi level, and it is likely that antiferro-orbital order exists in the electronic nematic phase. 
According to the previous ARPES study of BaFe$_2$As$_2$~\cite{myiPNAS}, broken $C_4$ rotational symmetry of the electronic structure persists up to $T\sim$ 170 K indicating the presence of \textit{ferro}-orbital order. 
Our results indicates that in addition to the ferro-orbital order, there also exists \textit{antiferro}-orbital ordering in the electronic nematic phase.

\section*{Acknowledgment}

We are grateful to H.~Kontani, S.~Onari, and Y.~Yamakawa for valuable discussion. We thank D. Hirai for his help in resistivity measurements. Transport measurements were also performed using facilities at the Cryogenic Research Center, the University of Tokyo. The experiment at Photon Factory was approved by the Photon Factory Program Advisory Committee (Proposals No. 2015S2-003).
This work was partly supported by the National Science and Technology Council of Taiwan under Grant No. 113-2112-M-007-033 and by the Japan Society for the Promotion of Science under Grant No. JP22K03535. 
A.F. acknowledges the support of the Yushan Fellow Program and the Center for Quantum Science and Technology within the framework of the Higher Education Sprout Project under the Ministry of Education of Taiwan.

\section*{Data Availability}

All data generated or analyzed during this study are available from the corresponding author upon reasonable request.

\bibliography{ref}

@article{sefat,
  title = {Superconductivity at 22 {K} in {C}o-Doped {B}a{F}e$_2${A}s$_2$ Crystals},
  author = {Sefat, Athena S. and Jin, Rongying and McGuire, Michael A. and Sales, Brian C. and Singh, David J. and Mandrus, David},
  journal = {Phys. Rev. Lett.},
  volume = {101},
  issue = {11},
  pages = {117004},
  numpages = {4},
  year = {2008},
  month = {Sep},
  publisher = {American Physical Society},
  doi = {10.1103/PhysRevLett.101.117004},
  url = {https://link.aps.org/doi/10.1103/PhysRevLett.101.117004}
}

@article{ljli,
  title = {Superconductivity induced by {N}i doping in {B}a{F}e$_2${A}s$_2$ single crystals},
  author = {L. J. Li and Y. K. Luo and Q. B. Wang and H. Chen and Z. Ren and Q. Tao and Y. K. Li and X. Lin and M. He and Z. W. Zhu},
  journal = {New. J. Phys.},
  volume = {11},
  pages = {025008},
  numpages = {4},
  year = {2009},
  publisher = {IOP Publishing Ltd.},
  doi = {10.1088/1367-2630/11/2/025008}
  }

@article{canfield,
  title = {Decoupling of the superconducting and magnetic/structural phase transitions in electron-doped {B}a{F}e$_2${A}s$_2$},
  author = {Canfield, P. C. and Bud'ko, S. L. and Ni, Ni and Yan, J. Q. and Kracher, A.},
  journal = {Phys. Rev. B},
  volume = {80},
  issue = {6},
  pages = {060501},
  year = {2009},
  month = {Aug},
  publisher = {American Physical Society},
  doi = {10.1103/PhysRevB.80.060501},
  url = {https://link.aps.org/doi/10.1103/PhysRevB.80.060501}
}

@article{sjiang,
  title = {Superconductivity up to 30 {K} in the vicinity of the quantum critical point in {B}a{F}e$_2$({A}s$_{1-x}${P}$_x$)$_2$},
  author = {Shuai Jiang and Hui Xing and Guofang Xuan and Cao Wang and Zhi Ren and Chunmu Feng and Jianhui Dai and Zhu'an Xu and Guanghan Cao},
  journal = {J. Phys.: Condens. Matter},
  volume = {21},
  pages = {382203},
  year = {2009},
  publisher = {IOP Publishing Ltd.},
  doi = {10.1088/0953-8984/21/38/382203}
  }

@article{nni,
  title = {Temperature versus doping phase diagrams for {B}a({F}e$_{1-x}${TM}$_x$)$_2${A}s$_2$ ({TM}$=${N}i, {C}u/{C}o)) single crystals},
  author = {Ni, N. and Thaler, A. and Yan, J. Q. and Kracher, A. and Colombier, E. and Bud'ko, S. L. and Canfield, P. C. and Hannahs, S. T.},
  journal = {Phys. Rev. B},
  volume = {82},
  issue = {2},
  pages = {024519},
  numpages = {16},
  year = {2010},
  month = {Jul},
  publisher = {American Physical Society},
  doi = {10.1103/PhysRevB.82.024519},
  url = {https://link.aps.org/doi/10.1103/PhysRevB.82.024519}
}

@article{qhuang,
  title = {Neutron-Diffraction Measurements of Magnetic Order and a Structural Transition in the Parent {B}a{F}e$_{2}${A}s$_{2}$ Compound of {F}e{A}s-Based High-Temperature Superconductors},
  author = {Huang, Q. and Qiu, Y. and Bao, Wei and Green, M. A. and Lynn, J. W. and Gasparovic, Y. C. and Wu, T. and Wu, G. and Chen, X. H.},
  journal = {Phys. Rev. Lett.},
  volume = {101},
  issue = {25},
  pages = {257003},
  numpages = {4},
  year = {2008},
  month = {Dec},
  publisher = {American Physical Society},
  doi = {10.1103/PhysRevLett.101.257003},
  url = {https://link.aps.org/doi/10.1103/PhysRevLett.101.257003}
}

@article{myiPRX,
  title = {Nematic Energy Scale and the Missing Electron Pocket in {F}e{S}e},
  author = {Yi, M. and Pfau, H. and Zhang, Y. and He, Y. and Wu, H. and Chen, T. and Ye, Z. R. and Hashimoto, M. and Yu, R. and Si, Q. and Lee, D.-H. and Dai, Pengcheng and Shen, Z.-X. and Lu, D. H. and Birgeneau, R. J.},
  journal = {Phys. Rev. X},
  volume = {9},
  issue = {4},
  pages = {041049},
  numpages = {12},
  year = {2019},
  month = {Dec},
  publisher = {American Physical Society},
  doi = {10.1103/PhysRevX.9.041049},
  url = {https://link.aps.org/doi/10.1103/PhysRevX.9.041049}
}

@article{kondo,
  title = {Unexpected Fermi-surface nesting in the pnictide parent compounds {B}a{F}e$_2${A}s$_2$ and {C}a{F}e$_2${A}s$_2$ revealed by angle-resolved photoemission spectroscopy},
  author = {Kondo, Takeshi and Fernandes, R. M. and Khasanov, R. and Liu, Chang and Palczewski, A. D. and Ni, Ni and Shi, M. and Bostwick, A. and Rotenberg, E. and Schmalian, J. and Bud'ko, S. L. and Canfield, P. C. and Kaminski, A.},
  journal = {Phys. Rev. B},
  volume = {81},
  issue = {6},
  pages = {060507},
  numpages = {4},
  year = {2010},
  month = {Feb},
  publisher = {American Physical Society},
  doi = {10.1103/PhysRevB.81.060507},
  url = {https://link.aps.org/doi/10.1103/PhysRevB.81.060507}
}

@article{lliu,
  title = {In-plane electronic anisotropy in the antiferromagnetic orthorhombic phase of isovalent-substituted {B}a({F}e$_{1-x}${R}u$_x$)$_2${A}s$_2$},
  author = {Liu, L. and Mikami, T. and Ishida, S. and Koshiishi, K. and Okazaki, K. and Yoshida, T. and Suzuki, H. and Horio, M. and Ambolode, L. C. C. and Xu, J. and Kumigashira, H. and Ono, K. and Nakajima, M. and Kihou, K. and Lee, C. H. and Iyo, A. and Eisaki, H. and Kakeshita, T. and Uchida, S. and Fujimori, A.},
  journal = {Phys. Rev. B},
  volume = {92},
  issue = {9},
  pages = {094503},
  numpages = {5},
  year = {2015},
  month = {Sep},
  publisher = {American Physical Society},
  doi = {10.1103/PhysRevB.92.094503},
  url = {https://link.aps.org/doi/10.1103/PhysRevB.92.094503}
}

@article{jhchu,
author= {J.-H. Chu and J. G. Analytis and K. De Greve and P. L. McMahon and Z. Islam and Y. Yamamoto and I. R. Fisher},
title ={In-Plane Resistivity Anisotropy in an Underdoped Iron Arsenide Superconductor},
journal = {Science},
volume = {329}, 
pages = {824}, 
year = {2010}, 
doi= {10.1126/science.1190482}
}

@article{ishidaPRB,
  title = {Manifestations of multiple-carrier charge transport in the magnetostructurally ordered phase of {B}a{F}e$_{2}${A}s$_{2}$},
  author = {Ishida, S. and Liang, T. and Nakajima, M. and Kihou, K. and Lee, C. H. and Iyo, A. and Eisaki, H. and Kakeshita, T. and Kida, T. and Hagiwara, M. and Tomioka, Y. and Ito, T. and Uchida, S.},
  journal = {Phys. Rev. B},
  volume = {84},
  issue = {18},
  pages = {184514},
  numpages = {7},
  year = {2011},
  month = {Nov},
  publisher = {American Physical Society},
  doi = {10.1103/PhysRevB.84.184514},
  url = {https://link.aps.org/doi/10.1103/PhysRevB.84.184514}
}

@article{nakajima,
  title = {Unprecedented anisotropic metallic state in undoped iron arsenide {B}a{F}e$_2$As$_2$ revealed by optical spectroscopy},
  author = {M. Nakajima and T. Liang and S. Ishida and Y. Tomioka K. Kihou and C. H. Lee and A. Iyo and H. Eisaki and T. Kakeshita and T. Ito and S. Uchida},
  journal = {Proc. Natl. Acad. Sci. U. S. A.},
  volume = {108},
  pages = {12238},
  year = {2011},
  doi = {10.1073/pnas.1100102108},
  url = {https://doi.org/10.1073/pnas.1100102108}
}

@article{dusza,
  title = {Anisotropic charge dynamics in detwinned {B}a({F}e$_{1-x}${C}o$_x$)$_2${A}s$_2$},
  author = {A. Dusza and A. Lucarelli and F. Pfuner and J.-H. Chu and I. R. Fisher and L. Degiorgi},
  journal = {Europhys. Lett.},
  volume = {93},
  pages = {37002},
  year = {2011},
  publisher = {IOP Publishing Ltd.},
  doi = {DOI 10.1209/0295-5075/93/37002}
  }

@article{jzhao,
author= {J. Zhao and D. T. Adroja and D.-X. Yao and R. Bewley and S. Li and X. F. Wang and G. Wu and X. H. Chen and J. Hu and P. Dai},
title ={Spin waves and magnetic exchange interactions in {C}a{F}e$_2${A}s$_2$},
journal = {Nat. Phys.},
volume = {5}, 
pages = {555}, 
year = {2009}, 
doi= {10.1038/nphys1336}
}

@article{myiPNAS,
  title = {Symmetry-breaking orbital anisotropy observed for detwinned {B}a({F}e$_{1-x}${C}o$_x$)$_2${A}s$_2$ above the spin density wave transition},
  author = {Ming Yi and Donghui Lu and Jiun-Haw Chu and James G. Analytis and Adam P. Sorini and Alexander F. Kemper and Brian Moritz and Sung-Kwan Mo and Rob G. Moore and Makoto Hashimoto and Wei-Sheng Lee and Zahid Hussain and Thomas P. Devereaux and Ian R. Fisher and Zhi-Xun Shen},
  journal = {Proc. Natl. Acad. Sci. U. S. A.},
  volume = {108},
  pages = {6878},
  year = {2011},
  doi = {10.1073/pnas.1100102108},
  url = {https://doi.org/10.1073/pnas.1100102108}
}

@article{ishidaPRL,
  title = {Anisotropy of the In-Plane Resistivity of Underdoped {B}a({F}e$_{1-x}${C}o$_x$)$_2${A}s$_2$ Superconductors Induced by Impurity Scattering in the Antiferromagnetic Orthorhombic Phase},
  author = {Ishida, S. and Nakajima, M. and Liang, T. and Kihou, K. and Lee, C. H. and Iyo, A. and Eisaki, H. and Kakeshita, T. and Tomioka, Y. and Ito, T. and Uchida, S.},
  journal = {Phys. Rev. Lett.},
  volume = {110},
  issue = {20},
  pages = {207001},
  numpages = {5},
  year = {2013},
  month = {May},
  publisher = {American Physical Society},
  doi = {10.1103/PhysRevLett.110.207001},
  url = {https://link.aps.org/doi/10.1103/PhysRevLett.110.207001}
}

@article{ishidaJACS,
  title = {Effect of Doping on the Magnetostructural Ordered Phase of Iron Arsenides: A Comparative Study of the Resistivity Anisotropy in Doped {B}a{F}e$_2${A}s$_2$ with Doping into Three Different Sites},
  author = {S. Ishida and M. Nakajima and T. Liang and K. Kihou and C. H. Lee and A. Iyo and H, Eisaki and T. Kakeshita and Y. Tomioka and T. Ito and S. Uchida},
  journal = {J. Am. Chem. Soc.},
  volume = {135},
  pages = {3158},
  year = {2013},
  publisher = {ACS Publication},
  doi = {10.1021/ja311174e},
  url = {https://doi.org/10.1021/ja311174e}
}

@article{kasahara,
author= {S.	Kasahara and H.	J. Shi and K. Hashimoto and S. Tonegawa and Y. Mizukami and T. Shibauchi and K. Sugimoto and T. Fukuda and T. Terashima and A. H. Nevidomskyy and Y. Matsuda},
title ={Electronic nematicity above the structural and superconducting transition in {B}a{F}e$_2$({A}s$_{1-x}${P}$_x$)$_2$},
journal = {Nature},
volume = {486}, 
pages = {555}, 
year = {2012}, 
doi= {10.1038/nature11178}
}

@article{achker,
author= {A. J. Achkar and M. Zwiebler and C. McMahon and F. He and R. Sutarto and I. Djianto and Z. Hao and M. J. P. Gingras and M. Hucker and G. D. Gu and A. Revcolevschi and H. Zhang and Y.-J. Kim and J. Geck and D. G. Hawthorn},
title ={Nematicity in stripe-ordered cuprates probed via resonant x-ray scattering},
journal = {Science},
volume = {351}, 
pages = {576}, 
year = {2016}, 
doi= {10.1126/science.aad1824}
}

@article{ronning,
author= {F. Ronning and T. Helm and K. R. Shirer and M. D. Bachmann and L. Balicas and M. K. Chan and B. J. Ramshaw and R. D. McDonald and F. F. Balakirev and M. Jaime and E. D. Bauer and P. J. W. Moll},
title ={Electronic in-plane symmetry breaking at field-tuned quantum criticality in {C}e{R}h{I}n$_5$},
journal = {Nature},
volume = {548}, 
pages = {313}, 
year = {2017}, 
doi= {10.1038/nature23315}
}

@article{shimojima,
  title = {Lifting of $xz/yz$ orbital degeneracy at the structural transition in detwinned FeSe},
  author = {Shimojima, T. and Suzuki, Y. and Sonobe, T. and Nakamura, A. and Sakano, M. and Omachi, J. and Yoshioka, K. and Kuwata-Gonokami, M. and Ono, K. and Kumigashira, H. and B\"ohmer, A. E. and Hardy, F. and Wolf, T. and Meingast, C. and L\"ohneysen, H. v. and Ikeda, H. and Ishizaka, K.},
  journal = {Phys. Rev. B},
  volume = {90},
  issue = {12},
  pages = {121111},
  numpages = {5},
  year = {2014},
  month = {Sep},
  publisher = {American Physical Society},
  doi = {10.1103/PhysRevB.90.121111}}

@article{cclee,
  title = {Ferro-Orbital Order and Strong Magnetic Anisotropy in the Parent Compounds of Iron-Pnictide Superconductors},
  author = {Lee, Chi-Cheng and Yin, Wei-Guo and Ku, Wei},
  journal = {Phys. Rev. Lett.},
  volume = {103},
  issue = {26},
  pages = {267001},
  numpages = {4},
  year = {2009},
  month = {Dec},
  publisher = {American Physical Society},
  doi = {10.1103/PhysRevLett.103.267001}
}

@article{yoshizawa,
  title = {Structural Quantum Criticality and Superconductivity in Iron-Based Superconductor {B}a({F}e$_{1-x}${C}o$_x$)$_2${A}s$_2$},
  author = {M. Yoshizawa and D. Kimura and T. Chiba and S. Simayi and Y. Nakanishi and K. Kihou and C. H. Lee and A. Iyo and H. Eisaki and M. Nakajima and S.-i. Uchida},
  journal = {J. Phys. Soc. Jpn.},
  volume = {81},
  pages = {024604},
  year = {2012},
  doi = {10.1143/JPSJ.81.024604}
}

@article{kontani,
  title = {Origin of orthorhombic transition, magnetic transition, and shear-modulus softening in iron pnictide superconductors: Analysis based on the orbital fluctuations theory},
  author = {Kontani, Hiroshi and Saito, Tetsuro and Onari, Seiichiro},
  journal = {Phys. Rev. B},
  volume = {84},
  issue = {2},
  pages = {024528},
  numpages = {17},
  year = {2011},
  month = {Jul},
  publisher = {American Physical Society},
  doi = {10.1103/PhysRevB.84.024528},
  url = {https://link.aps.org/doi/10.1103/PhysRevB.84.024528}
}

@article{ykim,
  title = {Electronic structure of detwinned {BaFe}$_{2}${As}$_{2}$ from photoemission and first principles},
  author = {Kim, Yeongkwan and Oh, Hyungju and Kim, Chul and Song, Dongjoon and Jung, Wonsig and Kim, Beomyoung and Choi, Hyoung Joon and Kim, Changyoung and Lee, Bumsung and Khim, Seunghyun and Kim, Hyungjoon and Kim, Keehoon and Hong, Jongbeom and Kwon, Yongseung},
  journal = {Phys. Rev. B},
  volume = {83},
  issue = {6},
  pages = {064509},
  numpages = {5},
  year = {2011},
  month = {Feb},
  publisher = {American Physical Society},
  doi = {10.1103/PhysRevB.83.064509}
}

@book{blaha,
author = {P. Blaha and K. Schwarz and G. K. H. Madsen and D. Kvasnicka and J. Luitz},
title = {WIEN2k, An Augmented Plane Wave + Local Orbitals Program for Calculating Crystal Properties}, 
publisher = {Technische Universitat Wien}, 
year = {2001},
address = {Vienna, Austria},
isbn = {ISBN 39501031-1-2}
}

@article{rotter,
  title = {Spin-density-wave anomaly at 140 {K} in the ternary iron arsenide {B}a{F}e$_2${A}s$_2$},
  author = {Rotter, Marianne and Tegel, Marcus and Johrendt, Dirk and Schellenberg, Inga and Hermes, Wilfried and P\"ottgen, Rainer},
  journal = {Phys. Rev. B},
  volume = {78},
  issue = {2},
  pages = {020503},
  numpages = {4},
  year = {2008},
  month = {Jul},
  publisher = {American Physical Society},
  doi = {10.1103/PhysRevB.78.020503}
}

@article{richard,
  title = {Observation of Dirac Cone Electronic Dispersion in {BaFe}$_{2}${As}$_{2}$},
  author = {Richard, P. and Nakayama, K. and Sato, T. and Neupane, M. and Xu, Y.-M. and Bowen, J. H. and Chen, G. F. and Luo, J. L. and Wang, N. L. and Dai, X. and Fang, Z. and Ding, H. and Takahashi, T.},
  journal = {Phys. Rev. Lett.},
  volume = {104},
  issue = {13},
  pages = {137001},
  numpages = {4},
  year = {2010},
  month = {Mar},
  publisher = {American Physical Society},
  doi = {10.1103/PhysRevLett.104.137001},
  url = {https://link.aps.org/doi/10.1103/PhysRevLett.104.137001}
}

@article{terashima,
  title = {Complete Fermi Surface in {BaFe}$_{2}${As}$_{2}$ Observed via {S}hubnikov--de {H}aas Oscillation Measurements on Detwinned Single Crystals},
  author = {Terashima, Taichi and Kurita, Nobuyuki and Tomita, Megumi and Kihou, Kunihiro and Lee, Chul-Ho and Tomioka, Yasuhide and Ito, Toshimitsu and Iyo, Akira and Eisaki, Hiroshi and Liang, Tian and Nakajima, Masamichi and Ishida, Shigeyuki and Uchida, Shin-ichi and Harima, Hisatomo and Uji, Shinya},
  journal = {Phys. Rev. Lett.},
  volume = {107},
  issue = {17},
  pages = {176402},
  numpages = {5},
  year = {2011},
  month = {Oct},
  publisher = {American Physical Society},
  doi = {10.1103/PhysRevLett.107.176402},
  url = {https://link.aps.org/doi/10.1103/PhysRevLett.107.176402}
}

@article{ishida-shibauchi,
  title = {Divergent Nematic Susceptibility near the Pseudogap Critical Point in a Cuprate Superconductor},
  author = {Kousuke Ishida and Suguru Hosoi and  Yuki Teramoto and Tomohiro Usui and Yuta Mizukami and Kenji Itaka and Yuji Matsuda and Takao Watanabe and Takasada Shibauchi},
  journal = {J. Phys. Soc. Jpn.},
  volume = {89},
  pages = {064707},
  year = {2020},
  doi = {10.1103/PhysRevLett.107.176402},
  url = {https://doi.org/10.7566/JPSJ.89.064707}
}

@article{nakata,
  title = {Nematicity in a cuprate superconductor revealed by angle-resolved photoemission spectroscopy under uniaxial strain},
  author = {S. Nakata and M. Horio and K. Koshiishi and K. Hagiwara and C. Lin and M. Suzuki and S. Ideta and K. Tanaka and D. Song and Y. Yoshida and H. Eisaki and A. Fujimori},
  journal = {npj Quantum Mater.},
  volume = {6},
  pages = {86},
  year = {2021},
  url = {https://doi.org/10.1038/s41535-021-00390-x}
}

@article{ren-yli,
  title = {Nematic Crossover in {BaFe}$_2${As}$_2$ under Uniaxial Stress},
  author = {Ren, Xiao and Duan, Lian and Hu, Yuwen and Li, Jiarui and Zhang, Rui and Luo, Huiqian and Dai, Pengcheng and Li, Yuan},
  journal = {Phys. Rev. Lett.},
  volume = {115},
  issue = {19},
  pages = {197002},
  numpages = {5},
  year = {2015},
  month = {Nov},
  publisher = {American Physical Society},
  doi = {10.1103/PhysRevLett.115.197002},
  url = {https://link.aps.org/doi/10.1103/PhysRevLett.115.197002}
}

@article{xlu,
author = {Xingye Lu  and J. T. Park  and Rui Zhang  and Huiqian Luo  and Andriy H. Nevidomskyy  and Qimiao Si  and Pengcheng Dai },
title = {Nematic spin correlations in the tetragonal state of uniaxial-strained {BaFe}$_{2-x}${Ni}$_x${As}$_2$},
journal = {Science},
volume = {345},
number = {6197},
pages = {657-660},
year = {2014},
doi = {10.1126/science.1251853},
URL = {https://www.science.org/doi/abs/10.1126/science.1251853}
}

@article{man-dai,
  title = {Electronic nematic correlations in the stress-free tetragonal state of {BaFe}$_{2-x}${Ni}$_x${As}$_2$},
  author = {Man, Haoran and Lu, Xingye and Chen, Justin S. and Zhang, Rui and Zhang, Wenliang and Luo, Huiqian and Kulda, J. and Ivanov, A. and Keller, T. and Morosan, Emilia and Si, Qimiao and Dai, Pengcheng},
  journal = {Phys. Rev. B},
  volume = {92},
  issue = {13},
  pages = {134521},
  numpages = {9},
  year = {2015},
  month = {Oct},
  publisher = {American Physical Society},
  doi = {10.1103/PhysRevB.92.134521},
  url = {https://link.aps.org/doi/10.1103/PhysRevB.92.134521}
}

\end{document}